\documentstyle[floats,prd,aps,epsfig]{revtex}
\input epsf 
\newcommand{\bce}{\begin{center}} \newcommand{\ece}{\end{center}}
\newcommand{\beq}{\begin{equation}} \newcommand{\eeq}{\end{equation}}
\newcommand{\beqy}{\begin{eqnarray}}
\newcommand{\eeqy}{\end{eqnarray}} \input epsf
\advance\voffset by 0.5in

\advance\voffset by -0.5in
\begin{document}
%\twocolumn [ \hsize\textwidth\columnwidth\hsize\csname @twocolumnfalse\endcsname
%
\title{Bremsstrahlung photons from the bare surface\\ of a strange quark star}
%age
\author{Prashanth Jaikumar\footnote{jaikumar@hep.physics.mcgill.ca}$^1$, Charles Gale$^1$, Dany Page$^2$, 
        and Madappa Prakash$^3$}
\address{$^1$Physics Department, McGill University, Montr\'{e}al,
         Qu\'{e}bec H3A 2T8, Canada.\\
 	 $^2$Instituto di Astronomia, UNAM, Mexico D.F. 04510, Mexico.
         \\              
         $^3$Department of Physics \& Astronomy, SUNY at Stony Brook,
        Stony Brook, NY 11794, USA.}
%\date{\today}
\maketitle
\begin{abstract}

The photon emissivity from the bremsstrahlung process
$e^-e^-\rightarrow e^-e^-\gamma$ occuring in the electrosphere at the
bare surface of a strange quark star is calculated. For surface
temperatures $T< 10^{9}$K, the photon flux exceeds that of $e^+e^-$
pairs that are produced via the Schwinger mechanism in the presence
of a strong electric field that binds electrons to the surface of the
quark star. The average energy of photons emitted from the
bremsstrahlung process can be 0.5 MeV or more, which is larger than
that in $e^+e^-$ pair annihilation. The observation of this
distinctive photon spectrum would constitute an unmistakable signature
of a strange quark star and shed light on color superconductivity at
stellar densities.


\end{abstract}
%------------------------------END OF TITLE AND ABSTRACT--------------
%\newpage
%\tableofcontents
%\newpage
%-----------------------------BEGIN INTRODUCTION-------------------

\section{Introduction}
\label{sec_intro}

Recently, it has been pointed out that thermal emission from the bare
surface of a strange quark star, due to both photons and $e^+e^-$ pair
production, can produce luminosities well above the Eddington limit
($\sim 10^{38}~{\rm erg~sec}^{-1}$) for extended periods of time, from
about a day to decades, depending on the superconducting phase of
quark matter~\cite{Page02}. The spectrum of emitted photons is
significantly different from that of a normal cooling neutron star
($30 <E/{\rm keV} < 500$ instead of $0.1 < E/{\rm keV} <2.5$).  This
distinctive spectrum and temperature evolution, if observed, would
constitute an unmistakable detection of a strange quark star and shed
light on color superconductivity at stellar densities. The predicted
characteristics are well within the capabilities of the INTEGRAL
satellite~\cite{Integral} launched toward the end of 2002.
\vskip 0.2cm
A strange quark matter (SQM) star  differs from a normal neutron star
(NS) in several important ways, which include

\vskip 0.2cm 

\noindent (1) A SQM star is
self-bound~\cite{witten84,Chin79,Farhi84,Haensel86,AO88} while a NS
requires gravitational forces for its binding.  As a consequence, the
pressure vanishes at vanishing baryon density in a NS with a surface
of normal matter.  The interior of the NS, however, may contain any or
a combination of non-nucleonic matter such as hyperons, pion or kaon
condensates, or quark matter~\cite{LP2003,Alford2001}. In contrast, a
self-bound SQM star is made entirely of quark matter up to its bare
surface and the pressure vanishes at a finite but supra-nuclear baryon
density. In the context of the MIT bag model with first order
corrections due to gluon exchange, the baryon density $n_B$ at which
the pressure vanishes is~\cite{PBP90}
\beq
\label{nzero}
n_B(P=0)=\biggl(\frac{4B}{3\pi^{2/3}}\biggr)^{3/4}
\biggl(1-\frac{2\alpha_c}{\pi}\biggr)^{1/4},
\eeq
where $B$ is the bag constant and $\alpha_c=g_c^2/(4\pi)$ is the
quark-gluon coupling constant. For typical values of $B$ and
$\alpha_c$~\cite{Farhi84}, the baryon density at vanishing pressure is about
2 to 3$n_0$, where $n_0=0.16$~fm$^{-3}$ is the saturation density of
nuclear matter. The density in Eq.~(\ref{nzero}) is not significantly
affected by the finite strange quark mass~\cite{PBP90} or by pairing
gaps in the quark phase~\cite{Reddy2002}. 

\vskip 0.1cm

\noindent (2) The mass versus radius relations of NS and SQM stars
differ significantly, although in the range of masses
($1<M/M_{\odot}<2$) observed to date \cite{TC99}, the calculated radii
are similar ($R\sim 10$ to 15 km). This makes it difficult to
distinguish these two classes on the basis of their gross physical
properties alone.

\vskip 0.2cm 

The light curves of NS and SQM stars, determined by surface photon
emission, can however be very different under certain conditions.
Central to the discussion of photon emission from a SQM star is the
physical size and composition of its surface layer. If quark matter is
in the Color-Flavor-Locked (CFL) superconducting phase (in which all
three flavors take part in pairing) up to the surface, no electrons
are required or admitted~\cite{RW03,AR02,SRP03}.  Surface photon
emission from the CFL phase has been found to saturate the blackbody
limit at early times~\cite{RVO}. In contrast, charge neutrality cannot
be satisfied by three flavors of quarks alone in the two-flavor
superconducting phase (2SC). In this case, an electron concentration
$n_e/n_B$ of about $10^{-4}$ to $10^{-3}$ is required to achieve
charge neutrality.  Normally, the bare surface of a quark star is an
inefficient emitter of thermal radiation below temperatures $T\sim
10^{11}$K because the plasma frequency of quark matter is high, of order
20 MeV~\cite{Alcock86}.  However, in the 2SC phase, strong electric
fields can arise because electrons in the surface layer are bound by
electrostatic interaction to quark matter. This layer of electrons,
which we will call the ``electrosphere'', is typically $10^{3}$ fm
thick~\cite{Alcock86} and can have radial electric fields whose
magnitude ($\sim 5\times 10^{17}~{\rm V~cm}^{-1}$) exceeds the
critical value required for electron-positron pair production from the
QED vacuum.  The critical value for the electric field was estimated
by Schwinger to be \cite{Schwinger}
\beq
E_{\rm cr}=\frac{m_e^2c^3}{e\hbar}\simeq 1.3\times 10^{16}~{\rm V~cm}^{-1}
\quad,
\eeq
where $m_e$ and $e$ are the electron's mass and charge,
respectively.  For a homogeneous electric 
field $E \gg E_{\rm cr}$, 
the vacuum pair-creation rate per unit volume is given by \cite{Schwinger}
\beq
R_{\pm}=\frac{m_e^4c^5}{24\pi\hbar^4}\biggl(\frac{E}{E_{\rm cr}}\biggr)^2\simeq 1.7\times 10^{50}\biggl(\frac{E}{E_{\rm cr}}\biggr)^2~{\rm cm}^{-3}~{\rm s}^{-1}
~.
\eeq
Usov~\cite{Usov98} first exploited the possibility of thermal emission
of hard photons from the annihilation of $e^+e^-$ pairs created by
super-critical electric fields in degenerate matter and
demonstrated that the associated photon emissivity is large enough to
be observable.  Subsequently, Page and Usov~\cite{Page02} have shown
that the bare surface of a quark star under super-critical electric
fields leads to a light curve that is significantly different from
those produced by either quark matter surrounded by a layer of normal
matter, or by surfaces of compact objects made entirely of normal
matter. Their principal conclusion was that hard photon emission from
$e^+e^-$ annihilation can dominate significantly over the blackbody
spectrum at lower luminosities, and that the variation in the photon
spectrum with luminosity could be used as an observational
signature of a young bare strange quark star. Furthermore,
calculations of light curves from quark matter in its superconducting
phases (CFL or 2SC) have shown that it may be possible to set bounds
on quark pairing gaps provided they are in the range 0.5 MeV to a few
MeV \cite{Page02}.

\vskip 0.2cm

Our objectives in this paper are to 

\noindent (1) point out that, depending on the temperature, the
bremsstrahlung process $e^-e^- \rightarrow e^-e^-\gamma$ in the
electrosphere can lead to photon luminosities that are significantly
larger than the $e^+e^-$ pair production luminosities in the presence
of strong electric fields, \\
\noindent (2) present a calculation of the corresponding emissivity 
including minimal screening effects, and 
investigate the role of the Landau-Pomeranchuk-Migdal (LPM)
effect, and \\
\noindent (3) compare the photon emissivity from the $e^-e^-
\rightarrow e^-e^-\gamma$ bremsstrahlung process to other processes
occurring in the thin electrosphere at the surface of the
star. Examples of such processes include $e^+e^-$ pair production
which occurs due to the presence of strong electric fields, and
equilibrium and non-thermal bremsstrahlung radiation from 
quark-quark collisions in the uppermost layer of quark matter. 

\vskip 0.2cm 

\noindent We wish to emphasize that our discussion applies only to the
case in which a charge neutralizing surface electron layer is present.
This can be realized in the 2SC, the gapless CFL (gCFL) or crystalline
color superconducting phases~\cite{AR02,KR}, but not in the pure CFL
phase.

\vskip 0.2cm

This paper is organized as follows. Sec. II is devoted to a
qualitative discussion of photon propagation in the
electrosphere. Here the physical characteristics of the electrosphere,
the photon mean free path, the photon dispersion relation, and the
role of the Landau-Pomeranchuk-Migdal (LPM) effect that suppresses the
emission of low-energy photons are addressed. The calculation of the
photon emissivity from the bremsstrahlung process in degenerate matter
(with and without modifications from the LPM effect) in the
electrosphere, and a characterization of the non-thermal nature of the
emitted photon spectrum are detailed in Sec. III.  This section also
contains numerical results and discussion of the bremsstrahlung
radiation flux, including a comparison with the flux of the
$e^+e^-$ pair annihilation process and quark-related processes.
Sec. IV contains our conclusions and outlook. Appendix A outlines the
evaluation of an integral required in Sec. III.

%-----------------------END OF INTRODUCTION-----------------------------------%
\vskip 0.2cm
\section{photon propagation in the electrosphere}
\noindent At the surface of the bare quark star, the density of quarks
drops abruptly to zero within a layer of thickness of about 1~fm, this
length scale being set by the range of the strong interaction. 
The density profile of the electrons can be determined by solving the
Poisson equation in the Thomas-Fermi approximation. In the
plane-parallel approximation for the layer of the electrosphere, the
electron chemical potential as a function of distance $z$ from the
quark surface is given by~\cite{NKG95}
\beq
\label{mue}
\mu_e(z)=\frac{\mu_e(0)}{1+z/H};\quad H=\frac{\hbar~c}{\mu_e(0)}
\sqrt{\frac{3\pi}{2\alpha}}=501.3
\biggl(\frac{10~{\rm MeV}}{\mu_e(0)}\biggr)~{\rm fm}\,,\label{eprofile}
\eeq
where $\alpha=e^2/4\pi$ is the fine structure constant. For a typical
$\mu_e(0)=(10-20)$ MeV, the electrosphere is about $10^3$~fm thick or
more.  The variation of $\mu_e$ across the electrosphere is shown in
Fig. 1.  The number density of electrons at the surface of such a star
is $n_e\sim 3\times(10^{-5}-10^{-4})~{\rm
fm}^{-3}$~\cite{Alcock86}. The electron degeneracy parameter
$\lambda=n_e^{1/3}h/\sqrt{m_eT}\gg 1$ for temperatures below
$10^{10}$ to $10^{11}$K ($\sim (1-10)$ MeV), which is the case some tens
of seconds after the star is born. Thereafter, electrons quickly
settle into a degenerate Fermi sea, and their Fermi momentum far
exceeds their rest mass, making it a relativistic degenerate
system. 
%Thus, in what follows, we will take the electron layer to be
%composed of a strongly degenerate ultra-relativistic ideal Fermi gas.

%
\begin{figure}[!ht]
\bce
\epsfig{file=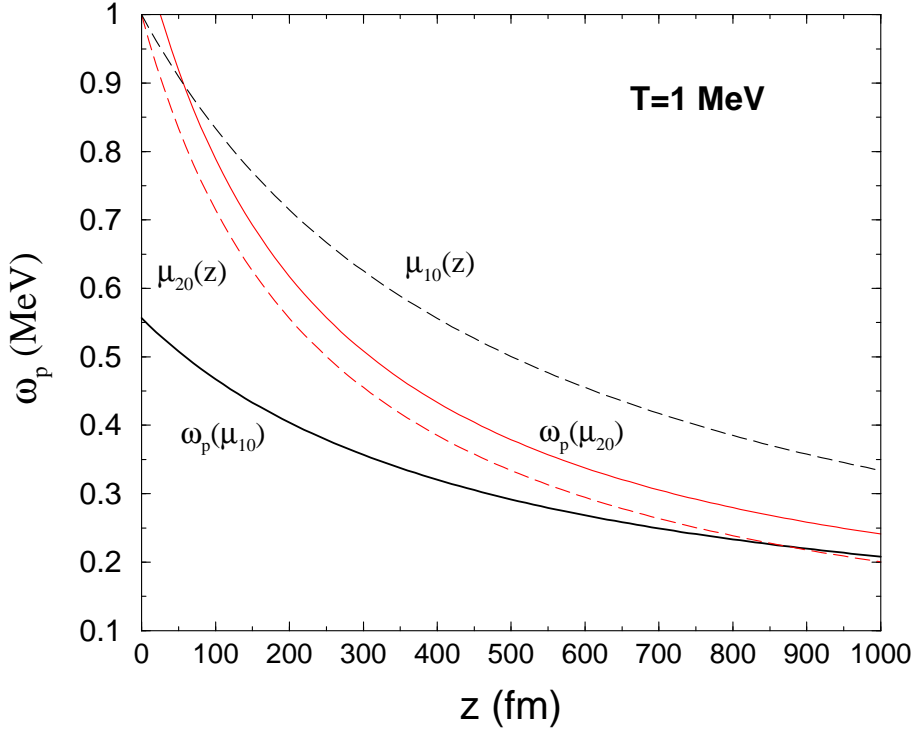,width=10.0cm,angle=270}
\ece
\caption{Profiles of the electron chemical potential $\mu_e$
(Eq.~(\ref{mue})) and the plasma frequency $\omega_p$ in the
degenerate limit (Eq.~(\ref{omegap})) versus distance $z$ from the
surface of the strange quark matter star. $\mu_{10}~(\mu_{20})$ is the
electron chemical potential in units of 10 (20) MeV. }
\label{figps1}
\end{figure} 
\vskip 0.2cm

The optical depth of the electron layer is determined by the mean
number of scatterings of photons on electrons. Bremsstrahlung photons
from this layer would mostly have
energies much smaller than the Fermi energy ($\epsilon_F=k_BT_F$) of the
electrons.  Since electrons can only manoeuvre within an energy window
$\Delta\epsilon\sim\frac{\pi^2}{12}\epsilon_F
\biggl(\frac{T}{T_F}\biggr)^2$ in a typical scattering process, and
$T\ll T_F$ due to strong degeneracy, the typical energy of the emitted
photons would be of order $\Delta\epsilon$.  The electron in the
intermediate state remains almost on-shell.

\vskip 0.1cm 

The ratio of the photon energy to electron mass is the small
parameter that justifies an estimate of the photon mean free path based on
the Thomson cross-section. If this ratio is of order unity or
larger, the Compton cross-section should be used. For a given electron
density, the photon mean free path is 
\beq
l_T\simeq\frac{1}{n_{eff}\sigma_T},
\quad \sigma_T=\frac{8\pi\alpha^2}{3m_e^2}=66.54~{\mbox {fm}}^2\quad,
\label{Thomson}
\eeq
where $\sigma_T$ is the Thomson cross section. The quantity $n_{eff}\approx
n_e(T=0)(T/\mu_e)$
is the effective number density of electrons, which takes into account
the fact that in a degenerate gas the available scattering states are 
restricted to the vicinity of the 
Fermi surface. For an electron density of $n_e=3.53\times
10^{-5}$~fm$^{-3}$ (corresponding to a chemical potential of
$\mu_e=20$ MeV), and $T=10^{11}$K (8.6 MeV), this yields a mean free
path $\sim 1000$ fm. For higher photon energies, the mean free path
using the Compton cross-section is somewhat larger.  The electron
layer, being about a 1000 fm thick, is therefore transparent to photons.

Since the emitted photons do not scatter often enough to become
thermalized, the electrosphere is a source of non-thermal low energy
photons.  We will quantify this non-thermal nature by calculating the
ratio of the second to first moment of the photon energy spectrum and
comparing it with a similar ratio (a fixed $T-$ independent number)
for the Planckian spectrum.

\vskip 0.2cm

To a very good approximation, the dispersion relation of the
emitted electromagnetic wave in the plasma can be taken as
$\omega=(\omega_p^2+k^2)^{1/2}$ (in $c=1$ units). The plasma frequency
$\omega_p$ serves as a low-energy cut-off. For degenerate electrons
($T/\mu_e \ll 1$), the plasma frequency is determined
from~\cite{SSP03}
\beq
\label{omegap}
\omega_p^2\cong\frac{4\alpha}{3\pi}
\mu_e^2\biggl(1+\frac{\pi^2T^2}{3\mu_e^2}\biggr) \,.
\eeq
Only photons with $\omega>\omega_p$ can propagate without severe
attenuation by the plasma. The plasma frequency, being 
density-dependent, 
decreases with increasing $z$ (Fig.~1). This implies that photons
emitted toward the surface from the lower layers of the electrosphere
can escape. A large plasma frequency, however, implies an overall
suppression of the emissivity. The optical depth of the electrosphere
at any distance is determined by the plasma frequency, which in turn
is a function of the electron density profile. The emissivity
$Q$ is thus a function of $\omega_p(z)$. The total
luminosity can be expressed as
\beq
L = 4\pi R_s^2\int_{z=0}^{z=z_0}Q(\omega_p(z))\,dz \,,
\eeq
where $z_0$ is the thickness of the electrosphere, and $R_s$ the radius of the star.

It should be noted that some (possibly paired) quark matter
is admixed with electrons in the innermost region (of thickness about
1~fm) of the electrosphere. The screening effect of quark or electron
matter will be taken into account in our calculation by modifying
appropriately the exchanged photon that mediates the $e^-e^-$
scattering. We will see that electric screening and magnetic damping
effects play only a small role for emission of low energy
photons. Such photons can escape despite the large plasma frequency
$\omega_p$ ($\sim 20$ MeV) of the ambient quark matter because the
thickness of the electron layer is about the same as $c/\omega_p\sim
10^{-12}$ cm.

\vskip 0.2cm

Bremsstrahlung radiation is known to be suppressed by
multiple scattering of electrons (LPM
effect~\cite{lpmigdal}) within the formation time of the emitted
photon. This is especially relevant for low energy photons (which
have a large formation time) that are emitted mostly in the
forward direction (small emission angles). The LPM effect is important
for wave numbers $k\leq k_c$ that satisfy~\cite{E146}
\beq
k_c=\frac{E_e^2}{E_{LPM}} \,,
\eeq
where $E_e$ is the typical energy of the emitting particle. For
degenerate electrons of energy $\mu_e\sim (10-20)$ MeV, the quantity
$E_{LPM}$ is given by~\cite{Drell96}
\beq
E_{LPM}({\rm eV})=3.8\times 10^{12}X_0({\rm cm}) \,.
\eeq
Above, $X_0$ is the typical radiation length in the medium. If, for
simplicity, we assume a static screened Coulomb interaction between
electrons, the radiation length is energy independent for high
energies of the emitting particles, and is given by~\footnote{Although
we use the zero-temperature expression for $n_e$ here, the available
scattering states are properly accounted for by the Fermi functions in
the LPM calculation to follow. }
\beq
X_0=\frac{m_e^2}{4n_e\alpha^3{\rm ln}(183)} \,,
\eeq
which yields $E_{LPM}=70$ keV and $k_c\sim 1.4$ GeV for 
$\mu_e=10$ MeV. This large value of $E_{LPM}$ implies that the LPM
effect will strongly suppress the number of photons radiated.  
For a photon of wave number $k$, the factor by which the
bremsstrahlung process is reduced is given by~\cite{E146}
\beq
S_{LPM}(k)=\sqrt{\frac{kE_{LPM}}{E_e^2}} \simeq \frac {1}{300}\,,
\eeq
where in quoting the numerical result above, we have used $\mu_e=10$~MeV
and $k=0.5~$MeV as an example.  This reduction shows that the
inclusion of the LPM effect is important for a realistic estimate of
the emissivity. The LPM suppression depends 
on the density of scatterers through $\mu_e$.
\vskip 0.2cm

\section{Photon Emissivity from the Bremsstrahlung process} 

The photon emissivity from the bremsstrahlung process is
\beqy
\label{emiss1} 
 Q = \frac{2\pi}{s\hbar}&&\biggl[\prod_{i=1}^{4}\int\frac{d^3p_i}{(2\pi)^32\omega_{p_i}}\biggr]\int\frac{d^3k}{(2\pi)^32\omega_k}\omega_k\sum_{spin}|M|^2~S_{LPM}(k)\\  \nonumber
&\times&n_F(\omega_{p_1})n_F(\omega_{p_2})
  \tilde{n}_F(\omega_{p_3})\tilde{n}_F(\omega_{p_4})(2\pi)^3\delta^3
({\bf P_f-P_i})\delta(E_f-E_i)\quad. 
\eeqy
The subscripts $i=1$ to 4 refer to electrons (1,2 label the incoming
states and 3,4 the outgoing states), whereas $(\omega_k,{\bf k})$
refer to the 4-momentum of the emitted photon. The symmetry factor
$s=2$. The phase space for electrons is convoluted with the
appropriate Fermi distribution functions $n_F(\omega_{p_i})=1/({\rm
e}^{\omega_{p_i}/T}+1)$ (in $k_B=1$ units) and ${\tilde n_F}=1-n_F$,
respectively.  For simplicity, and to highlight the role of the LPM
effect, we first obtain the analytic $T~{\rm and }~\mu_e$ dependences
of the emissivity without the LPM suppression factor. The effects of
the LPM effect are investigated in the second stage. In this case, an
analytical analysis becomes unwieldy and hence we present results of
numerical calculations. A comparison of the respective emissivities
enables an overall assessment of the importance of in-medium multiple
scattering effects.
\vskip 0.2cm 

\subsection{The Bremsstrahlung Process in Degenerate Matter }

For conditions of interest here, electrons in the electrosphere are
strongly degenerate. The emissivity will thus be dominated by low
energy photons ($\omega_p<\omega_k\ll\mu_e$).  The contribution from
high energy photons ($\omega_k\gg m_e$), which requires a full quantum
calculation of the bremsstrahlung process, is expected to be small.
We utilize the quasiclassical approximation for low energy photons, in
which case the calculation of the emissivity simplifies if we use Low's
theorem for bremsstrahlung~\cite{Low58}.

Low's theorem states that the first two terms in the differential
cross-section of bremsstrahlung, expressed as a power series in the
energy loss $\omega_k$, is unique and exactly calculable in terms of
the corresponding elastic amplitude. The presence of the factor
$\omega_k$ in the numerator of Eq.~(\ref{emiss1}) ensures that in the
low energy limit, the dominant contribution to $Q$ is non-singular and
can be estimated once the leading term in the series ($\sim
1/\omega_k$) is established. Retaining only the first term in Low's
soft-photon expansion shows that the full scattering amplitude for
bremsstrahlung factorizes into an elastic part, which is M$\o$ller
scattering, and a part which gives the classical intensity of photon
emission divided by the energy $\omega_k$. At low energy, the
probability of successive photon emission is enhanced, which causes a
naive perturbative expansion in powers of the electromagnetic coupling
to fail. The differential cross-section can alternatively be
understood in terms of the mean number of photons emitted, which is
equivalent to the classical intensity divided by the mean energy of
one photon. This factorization of amplitudes simplifies our
calculation to a large extent. The energy exchange in $e^-e^-$
scattering being small, we can write the matrix element explicitly as
\beq
iM=M_{el}\biggl[e\biggl(\frac{p_4.\epsilon^{\ast}}{p_4.k}-\frac{p_2.\epsilon^{\ast}}{p_2.k}+\frac{p_3.\epsilon^{\ast}}{p_3.k}-\frac{p_1.\epsilon^{\ast}}{p_1.k}\biggr)\biggr]\quad,\label{decouple}
\eeq
 where $M_{el}$ denotes the interaction for elastic scattering of the
incoming electrons, $\epsilon^{\mu}$ denotes the polarization vector
of the photon, and $e$ is the charge of the electron. The term in the
second bracket in Eq.~(\ref{decouple}) is the familiar result from
classical radiation theory for an electron that accelerates under an
impulsive force. In the low-energy limit, the emission from the
exchange diagrams simply doubles the overall result for the emissivity
from the direct diagrams (no interference). Since there is a symmetry
factor of $s=2$ in the denominator of the emissivity expression
Eq.~(\ref{emiss1}), we may ignore the identity of the particles
altogether and evaluate the emissivity from the direct diagrams
alone. The spin sum can be performed separately over the electrons and
the photon thanks to the decoupling at low energy. We are thus led to
the following expression for the emissivity
\beqy
\label{emiss22} 
Q&=&\frac{1}{\hbar}\int\frac{d^3p_1}{(2\pi)^3}\frac{d^3p_2}{(2\pi)^3}n_F(\omega_{p_1})n_F(\omega_{p_2})\int d\omega_k\,\omega_k\biggl(\frac{d\sigma}{d\omega_k}\biggr) \quad;\\ 
\biggl(\frac{d\sigma}{d\omega_k}\biggr)&=&(d\sigma_{el})\frac{\sqrt{\omega_k^2-\omega_p^2}}{2}\int\frac{d\Omega_k}{(2\pi)^3}\sum_{\lambda=1,2}e^2\left|\frac{p_4.\epsilon_{\lambda}^{\ast}}{p_4.k}-\frac{p_2.\epsilon_{\lambda}^{\ast}}{p_2.k}+\frac{p_3.\epsilon^{\ast}}{p_3.k}-\frac{p_1.\epsilon^{\ast}}{p_1.k}\right|^2 \label{factor}\quad,\\ \nonumber
(d\sigma_{el})&=&\frac{1}{2\omega_{p_1}}\frac{1}{2\omega_{p_2}}\biggl(\prod_{f=3,4}\frac{d^3p_f}{2\omega_{p_f}}\biggr)\sum_{s=\pm}|M_{el}|^2(2\pi)^3\delta^3({\bf P_f-P_i})2\pi\delta(E_f-E_i)\tilde{n}_F(\omega_{p_3})\tilde{n}_F(\omega_{p_4}) \quad.\nonumber
\eeqy

The change of variable from $k$ to $\omega_k$ introduces the square
root factor as the Jacobian, and the integration over $\omega_k$ runs
from $\omega_p$ to $\infty$. Medium modifications to bremsstrahlung
emission can be encapsulated in the refractive index of the electron
plasma $n=\sqrt{1-\omega_p^2/\omega_k^2}$. With the definitions ${\bf
p} = E {\bf v}$~{\rm and}~${\bf k} = \omega_k {\bf n} = \omega_k n
\hat{\bf n}$, where $|\hat{\bf n}|=1$, the angular integral in
Eq.~(\ref{factor}) is
\beq
\int\frac{d\Omega_k}{(2\pi)^3}\sum_{\lambda=1,2}\frac{e^2}{\omega_k^2}\left[\left(\frac{p_4\cdot\epsilon^*}{E_4\tilde{n}_-\left[1-\frac{n}{\tilde{n}_-} \;{\bf v}_4\cdot{\bf \hat{n}}\right]} 
-\frac{p_2\cdot\epsilon^*}{E_2\tilde{n}_+ 
\left[1-\frac{n}{\tilde{n}_+}\;{\bf v}_2\cdot{\bf\hat{n}}\right]} 
+\frac{p_3\cdot\epsilon^*}{E_3\tilde{n}_- 
\left[1-\frac{n}{\tilde{n}_-}\;{\bf v}_3\cdot{\bf \hat{n}}\right]} 
-\frac{p_1 \cdot \epsilon^*}{E_1\tilde{n}_+ 
\left[1-\frac{n}{\tilde{n}_+}\;{\bf v}_1\cdot{\bf \hat{n}}\right]} 
\right)\right]^2
\eeq
where $\tilde{n}_\pm  \equiv 1 \pm  \frac{1}{2}\frac{\omega_p^2}{E \omega_k}$
and we have neglected terms like $\gamma \cdot \epsilon^* \, \gamma
\cdot k$ in the numerators compared to $\epsilon^*\cdot p$. Since
$\omega_k\simeq\omega_p\ll E\simeq \mu_e$, we may approximate
$\hat{n}_+ = \hat{n}_- =1$. Although the refractive index $n\ll 1$, we
cannot neglect the factors $[1-{\bf v}_i\cdot{\bf \hat{n}}]$ in the
denominator, since that would make the matrix element vanish by
momentum conservation ($\sum_{i}v_{i}=0$).  Thus, for small $n$, the
leading contribution to the squared matrix element for bremsstrahlung
is ${\cal O}(n^2)$ and to the cross section ${\cal O}(n^3)$, one power
of $n$ coming from the Jacobian in Eq.~(\ref{factor}).   This allows
us to perform an expansion of the angular integral in even powers of
$n$. To ${\cal O}(n^2)$, and in the ultra-relativistic limit, the
angular integral over $d\Omega_k$ can be evaluated, and one finds 
\beq
\biggl(\frac{d\sigma}{d\omega_k}\biggr)=(d\sigma_{el})
\biggl[\frac{8\alpha}{5\pi}\frac{n^3{\rm
sin}^2\theta_{CM}}{\omega_k}\biggr]\,, 
\eeq 
where $\alpha=e^2/(4\pi)$ is the fine structure constant and
$\theta_{CM}$ is the angle in the centre of mass (henceforth CM)
frame. In comparison, the vacuum result is~\cite{LL}
\beqy
\biggl(\frac{d\sigma}{d\omega_k}\biggr)=(d\sigma_{el})\biggl[\alpha
\frac{F(\xi)}{\omega_k}\biggr]\,, 
\eeqy
with
\beqy
F(\xi) = \frac {2}{\pi} \left[ 
\frac {2\xi^2+1}{\xi{\sqrt{\xi^2+1} }} \ln (\xi + {\sqrt{(\xi^2+1)}})
- 1 \right]\,; \qquad \qquad \xi = 
\frac{E}{m_e}{\rm sin}\biggl(\frac{\theta_{CM}}{2}\biggr)\quad.
\nonumber
\eeqy
The function $F(\xi)$ reduces to $8\xi^2/3\pi$ in the limit $\xi\ll 1$
(small angle scattering in the CM frame) which is the region of
interest here. A comparison of the vacuum and medium expressions 
reveals a suppression factor of order $n^3\times (m_e/\mu_e)^2$ due
to the refractive index $n$ of the medium.

In general, there are two conditions that must be satisfied for the
validity of the two preceding equations. The first is that $m_e/E\ll
1$, because electromagnetic radiation from fast-moving particles is
emitted roughly in a very narrow cone of this opening angle (this is
equivalent to $\xi\ll 1$). Secondly, the exchanged photon 3-momentum
${\bf q}$ should be much larger than the change in ${\bf q}$ induced
by soft photon emission~\cite{LL}. Explicitly,
\beq
\biggl(\frac{\omega_km_e^2}{E^3}\biggr)\ll\theta_{CM}\ll
\biggl(\frac{m_e}{E}\biggr) \label{bounds}
\eeq  
in the ultra-relativistic case.  These limits on $\theta_{CM}$ 
suffice to give results for $d\sigma/d\omega_k$ that are correct to
logarithmic accuracy. To be more accurate, one needs to go beyond the
quasi-classical approximation which is not required for our purposes,
since we are only interested in the emission of low energy
photons. Emission of photons with large energies is severely supressed by Pauli
blocking in the degenerate electron sea, and makes a negligible
contribution to the emissivity.

\vskip 0.2cm

The soft photon limit also simplifies the calculation of the in-medium
elastic scattering cross section ($d\sigma_{el}$). The 
scattering occurs via the exchange of a screened photon
propagator. Based on the general principle of gauge invariance, we can
use Weldon's decomposition~\cite{Weldon82} of the amplitude 
\beq
M_{el}=J_{\mu}(p_1,p_3)D^{\mu\nu}(q_0,{\bf q})J_{\nu}(p_2,p_4);
\quad J_{\mu}(p_i^{in},p_j^{out})=e{\bar u}(p_j)\gamma_{\mu}u(p_i) \quad,
\eeq 
and write it in terms of the transverse and longitudinal (temporal)
parts of the one-loop self-energy as
\beq
M_{el}=J_{\mu}D^{\mu\nu}(q)J_{\nu}=\frac{J_0J_0}{{\bf q}^2+\Pi_L}+\frac{J_tJ_t}{q^2-\Pi_T}\quad,
\eeq
where the exchange 4-momentum $q=(q_0,{\bf
q})=p_1-p_3=p_4-p_2$. Heiselberg and Pethick~\cite{HP93} have shown
that in the limit of small $q_0,|{\bf q}| \ll E\sim \mu_e$, the magnetic
part of the electron current (spin flips) can be neglected, as is
evident from Gordon's identity~\cite{Peskin}. In this case, the
electron current reduces to $J_{\mu}(p_i^{in},p_j^{out})\rightarrow
2ep_{i_{\mu}}^{in}$. The squared matrix element then becomes
\beq
|M_{el}|^2=16~e^4E_1^2E_2^2\left|\frac{1}{{\bf q}^2+\Pi_L} 
+\frac{(1-x^2){\rm cos}~\phi}{q^2-\Pi_T}\right|^2\,, 
\label{ematrix}
\eeq
where $x=q_0/|{\bf q}|$ and $\phi$ is the angle between the components
of the velocities of the incoming particles transverse to the momentum
transfer ${\bf q}$. While $q_0$ is small, the momentum transfer ${\bf q}$ can be larger since the maximum angle by which electrons can scatter on the
Fermi surface is set by the upper bound in Eq.~(\ref{bounds}).

The longitudinal and transverse self-energies $\Pi_L$ and $\Pi_T$ are
in general complex functions of $q_0$ and ${|\bf q|}$. These
self-energies can be obtained in the hard dense loop
approximation~\footnote{We are ignoring the smaller contribution from
finite temperature effects (hard thermal loops).}.  Since the
exchanged photon also has small energy, we will use approximate
forms~\cite{Lebellac} in the nearly static limit $q_0/{|\bf q|}\ll 1$:
\beq
\Pi_L=m_D^2 \quad {\rm and} \quad 
\Pi_T=i\frac{\pi}{4}\frac{q_0}{{|\bf q|}}m_D^2\quad.
\eeq
For degenerate unpaired quark matter 
\beq 
m_D^2=(\sum z_i^2 e^2)~\mu_q^2 \,,
\eeq
where $z_i = +2/3$ or $-1/3$ is the charge of the unpaired quark
species $i$. 

In the 2SC phase, up and down quarks of one color are
unpaired. Strange quarks can be unpaired or form a spin one condensate
which breaks rotational invariance. For the paired quarks in
the 2SC phase, photon-gluon mixing can occur. The residual $U(1)$
symmetry is a combination of the electric charge $Q$ and the $SU(3)$
hypercharge $T_8$. The Debye mass associated with the photon is 
given by~\cite{Dirk03}
\beq
m_D^2=\frac{2e^2\mu_q^2}{3\pi^2} \,,
\eeq
where $\mu_q$ is the quark chemical potential, of order hundreds of
MeV. For degenerate electrons, 
\beq 
m_D^2=e^2\mu_e^2/3 \,.  
\eeq 
In what follows, we denote the phase dependent screening mass by
$m_D$, noting that its magnitude is generally larger than $m_e$.
Contributions to the emissivity from screened electric effects will
turn out to be much smaller than those from unscreened magnetostatic
modes.

\vskip 0.2cm

It is easier to evaluate $d\sigma_{el}$, a Lorentz
invariant quantity, in the CM frame, and later transform to laboratory
co-ordinates for the final phase space integrations. In the CM frame,
elastic scattering between identical particles implies that $q_0=0$
and ${\rm cos}~\phi=-1$. If the scattering angle $\theta_{lab}$ is
small, so is the scattering angle $\theta_{CM}$ in the CM frame 
since $\theta_{CM}=2\theta_{lab}$ for identical particles. Moreover,
$q^2=-{\bf q}^2=-E_1E_3\theta_{CM}^2$. The elastic differential
cross-section then simplifies to
\beqy
d\sigma_{el}&=&\frac{1}{2E_1}\frac{1}{2E_2}\int\frac{d^3p_3}{(2\pi)^32E_3}\frac{d^3p_4}{(2\pi)^32E_4}16e^4E_1^2E_2^2\left|\frac{1}{E_1E_3{\theta_{CM}}^2+m_D^2}+\frac{1}{E_1E_3{\theta_{CM}}^2}\right|^2\\ \nonumber
&&\times(2\pi)^3\delta^3({\bf P_1+P_2-P_3-P_4})2\pi\delta(E_1+E_2-E_3-E_4)\tilde{n}_F(E_3)\tilde{n}_F(E_4)\quad.
\eeqy
Denoting the total CM energy as $E_1+E_2=2{\cal E}$ and using the momentum conserving delta-function to perform the $d^3p_4$ integration, we find
\beq
d\sigma_{el}=2\alpha^2{\cal E}^2\int\frac{d^3p_3}{E_3^2}\left|\frac{1}{E_1E_3{\theta_{CM}}^2+m_D^2}+\frac{1}{E_1E_3{\theta_{CM}}^2}\right|^2\delta({\cal E}-E_3)(\tilde{n}_F(E_3))^2\quad,
\eeq
where $\alpha=e^2/4\pi$ is the fine structure constant. Since the main
contribution comes from the vicinity of
the Fermi surface, we can change variables as
$d^3p_3\approx E_3^2dE_3d\Omega_{p_3}$ (the mass of the electron can
be safely ignored here). The delta function facilitates the integration over
$d|{\bf p_3}|$ with the result
\beq
d\sigma_{el}=\frac{2\alpha^2}{{\cal E}^2}(\tilde{n}_F({\cal E}))^2\int d\Omega_{p_3}\biggl(\frac{1}{\theta_{CM}^2+a^2}+\frac{1}{\theta_{CM}^2}\biggr)^2;\quad a=m_D/{\cal E}\quad. \label{inmedium}
\eeq
Without loss of generality, we can take ${\bf P_1}$ and ${\bf P_2}$
to be along ${\bf \hat{z}}$ and ${-\bf \hat{z}}$, respectively, so that
$d\Omega_{p_3}=d\Omega_{CM}$. It can be checked at this stage that
we recover the vacuum result for M$\o$ller
scattering. If the blocking factors from the above expression are set
to unity, and $a=0$, we find
\beq
\frac{d\sigma_{el}}{d\Omega_{CM}} = 
\frac{4\alpha^2}{{\cal E}^2\theta_{CM}^4}\quad,
\eeq
which is the most singular part of the M$\o$ller cross-section in the
ultra-relativistic limit.  (In making the comparison, a symmetrizing
factor of 1/2 has to be reinserted, because particles in the initial
state are indistinguishable). Returning to the in-medium expression
Eq.~(\ref{inmedium}), we note that the integral over $d\theta_{CM}$
should also include the $\theta_{CM}$-dependence from $F(\xi)$.
Integrating $\phi$ from 0 to $2\pi$, and $\theta_{CM}$ from
$\theta_{min}$ to $\theta_{max}$ as prescribed by Eq.~(\ref{bounds}),
we obtain
\beq
d\sigma_{el}^{\prime} = 
\frac{4\pi\alpha^2}{{\cal E}^2}
(\tilde{n}_F({\cal E}))^2\biggl[{\rm ln}
\biggl(\frac{\theta_{max}}{\theta_{min}}\biggr)
+ \frac{3}{2}{\rm ln}
\biggl(\frac{\theta_{max}^2+a^2}{\theta_{min}^2+a^2}\biggr)
+ \frac{a^2}{2}
\biggl(\frac{1}{\theta_{max}^2+a^2}
- \frac{1}{\theta_{min}^2+a^2}\biggr)\biggr]\quad.
\eeq
The prime on $d\sigma_{el}$ is to remind us that the angle-dependent
part from photon emission must also be included in performing the
angular integral in the elastic cross-section. Next, the $d\omega_k$
integral in Eq.~(\ref{emiss22}) is to be performed. The lower limit on
$\omega_k$ is  $\omega_{min}=\omega_p$, while the upper limit
is  $\omega_{max}\approx \sqrt{\omega_p^2+m_e^2}$. The upper
limit is set by noting that for small angles, we can write ${\bf
|k|}_{max}\approx E\theta_{max}\approx E(m_e/E)=m_e$. Although
photons of energy $\omega_k>\omega_{max}$ are also emitted, their
contribution, which cannot be included consistently within our
approximations, is expected to be small in the degenerate limit.

\vskip 0.1cm

The integral over $\omega_k$ can be rewritten as
\beq
\label{spectrum}
\int
d\omega_k\,\omega_k\biggl(\frac{d\sigma}{d\omega_k}\biggr) = 
\frac{32\alpha^3}{5{\cal
E}^2}(\tilde{n}_F({\cal
E}))^2\int_{\omega_{min}}^{\omega_{max}}d\omega_k 
n^3
\biggl\{I_1+I_2+I_3\biggl\}\quad, \label{eyes}
\eeq
where
\beqy
I_1&=& {\rm ln}\biggl(\frac{E^2}{m_e\omega_k}\biggr)\quad,  \label{eye1}\\ 
I_2&=&\frac{3}{2}\biggl[{\rm ln}\biggl(\frac{m_e^2+m_D^2}{E^2}\biggr)-{\rm ln}\biggl(\biggl(\frac{m_e^2\omega_k}{E^3}\biggr)^2+\frac{m_D^2}{E^2}\biggr)\biggr]\quad,\\ 
I_3&=&\frac{m_D^2}{2(m_e^2+m_D^2)}-\frac{m_D^2E^4}{2m_e^4}\frac{1}{(\omega_k^2+(\frac{m_DE^2}{m_e^2})^2)}\quad.
\eeqy
The Debye mass $m_D\gg m_e$, which allows us to 
drop  contributions from $I_2$ and $I_3$, and retain the largest
contribution from $I_1$. This contribution is easy to
evaluate, since $\omega_p$ is a slowly varying function of
the distance $z$ for temperatures less than 1 MeV~\cite{NKG95}. The
logarithmic factor in Eq.~(\ref{eye1}) is only weakly dependent on
$\omega_k$ as compared to the square root prefactor and can therefore
be pulled out of the integral with only a slight underestimation of
the integral. This leads to the expression
\beq
\int \omega_k d\sigma=\frac{64\alpha^3m_e}{5{\cal E}^2}(\tilde{n}_F({\cal E}))^2{\rm ln}\biggl(\frac{E}{m_e}\biggr)\biggl\{1+\frac{1}{2}\frac{\omega_p^2}{\omega_p^2+m_e^2}-\frac{3}{2}{\rm tan}^{-1}\biggl(\frac{m_e}{\omega_p}\biggr)\biggr\} \label{Ione}\quad.
\eeq 
Inserting this result in Eq.~(\ref{emiss22}), the emissivity becomes 
\beq
Q=\frac{64\alpha^3m_e}{5\hbar}\int\frac{d^3p_1}{(2\pi)^3}\frac{d^3p_2}{(2\pi)^3}n_F(\omega_{p_1})n_F(\omega_{p_2})\frac{\tilde{n}_F({\cal
E})^2}{{\cal E}^2}{\rm ln}\biggl(\frac{E}{m_e}\biggr)
\biggl\{1+\frac{1}{2}\frac{\omega_p^2}{\omega_p^2+m_e^2}-\frac{3\omega_p}{2m_e}{\rm tan}^{-1}\biggl(\frac{m_e}{\omega_p}\biggr)\biggr\} 
\label{emiss3}\quad.
\eeq
We note that electric effects are not very important for the
conditions of interest here, since they are strongly supressed because 
$(m_e/m_D)^2\ll 1$. We have therefore dropped 
contributions of ${\cal O}\biggl(m_e^2/m_D^2\biggr)$. As we have
focused on soft exchanges, the magnetic damping for such modes
vanishes; consequently they provide the largest contribution to the
integral.

For strongly degenerate electrons, further reduction of
Eq.~(\ref{emiss3}) is afforded by the nearly step function character of 
the Fermi functions.  We can approximate $E\approx \mu_e$, 
so that the logarithmic factor is now a constant
and can be pulled out of the integral. The variable ${\cal E}$ can be
converted to laboratory coordinates using the relation
\beqy
{\cal E}&=&\biggl\{\frac{|\bf P_1||\bf P_2|}{2}(1-{\rm cos}~\theta_{12})+\frac{m_e^2}{4}\biggl(2+\frac{|\bf P_1|}{|\bf P_2|}+\frac{|\bf P_2|}{|\bf P_1|}\biggr)\biggr\}^{1/2}\quad,\\ \nonumber
&&{\rm cos}~\theta_{12}={\rm cos}~\theta_1{\rm cos}~\theta_2+{\rm sin}~\theta_1{\rm sin}~\theta_2{\rm cos}~(\phi_1-\phi_2)\quad.
\eeqy
We can further set $|{\bf P_1}|,|{\bf P_2}|\approx p_{F_e}$, the electron
Fermi momentum so that $\tilde{n}_F({\cal E})$ depends only on the
angular variables $\theta_1$ and $\theta_2$. The integral over the
momenta can be evaluated approximately in the degenerate limit:
\beq {\cal N}(T,\mu_e)=\int dp\frac{p^2}{{\rm
e}^{(\sqrt{p^2+m_e^2}-\mu_e)/T}+1}\longrightarrow
\frac{\mu_e^3}{3}\biggl[1+\biggl(\frac{\pi T}{\mu_e}\biggr)^2+{\cal
O}\biggl(\frac{m_e^4T^4}{\mu_e^8}\biggr)+...\biggr]\quad.  \eeq
Terms involving the electron mass can be dropped because they  
appear only beyond second order in the above expansion.
More details on the evaluation of the angular
integral are given in Appendix A. Here, we quote the final result for
the emissivity in two relevant limits:
\beqy 
\label{finale} 
Q&=&\frac{64\alpha^3m_e{\cal F}(\omega_p)({\cal
N}(T,\mu_e))^2}{5\hbar(2\pi)^6}{\rm
ln}\biggl(\frac{\mu_e}{m_e}\biggr){\cal
I}(T,\mu_e) \label{efinal}\\ \nonumber 
{\cal F}(\omega_p) &=&
\biggl\{1+\frac{1}{2}\frac{\omega_p^2}{\omega_p^2+m_e^2}-\frac{3\omega_p}{2m_e}{\rm tan}^{-1}\biggl(\frac{m_e}{\omega_p}\biggr)\biggr\} 
\\ \nonumber
{\cal I}(T,\mu_e) & = & 
8\pi^2\times\left\{\begin{array}{c}\frac{4T}{\mu_e^3}\biggl({\rm ln}~2-
\frac{1}{2}\biggr);\quad
\frac{m_e^2}{2\mu_e}\leq T \ll 
\mu_e \\ \frac{2}{\mu_e^2}{\rm e}^{-m_e^2/2\mu_e T};\quad
T\ll\frac{m_e^2}{2\mu_e}\end{array}\right\} \,.
\eeqy 
Note that ${\cal F}(\omega_p) \rightarrow 1$ for $\omega_p \ll m_e$
and ${\cal F}(\omega_p) \rightarrow (m_e/\omega_p)^4/5$ for $\omega_p \gg m_e$.

Equation (\ref{finale}) provides the photon emissivity to logarithmic
accuracy. A better estimate can be obtained by a numerical integration
of Eq.~(\ref{emiss3}) which can be reduced to a five-dimensional
integral by choosing, for example, the azimuthal angle $\phi_2=0$.
The limits on $\phi_1$ then run from 0 to $2\pi$, and $\theta_1$ and
$\theta_2$ each range from 0 to $\pi$.  In the degenerate limit, the
Fermi functions cause the integrand in Eq.~(\ref{emiss3}) to be
strongly peaked for radial momenta $p_1 \approx
p_2\approx\mu_e$.  In this case, it is advantageous to use an adaptive
Monte-carlo method such as VEGAS \cite{NR}, which is based on
stratified importance sampling.  We can restrict the range of
integration for the radial momenta from $\mu_e-\delta$ to $\mu_e+\delta$,
where $\delta\ll\mu_e$ is chosen to be a suitable multiple of the
temperature. This procedure provides an accuracy of better than a few
percent. 

If the temperature becomes comparable to $\mu_e$, electrons cannot be
considered as degenerate and our approximations would not be valid. In
our numerical calculations therefore, temperatures are chosen to be
much smaller than $\mu_e$.  Within the temperature limits that set the
validity of our expressions, one can compare the photon emissivity
from $e^-e^-$ bremsstrahlung to that of $e^+e^-$
pairs~\cite{Usov01}. Such comparisons are presented below.

\vskip 0.2cm

%\subsection{Comparisons of the Bremsstrahlung Radiation Intensity with Those
%from Other Sources}

\subsection{Comparison with Other Processes}

For purposes of comparison and easy estimation, it is useful to
express all fluxes in terms of typical temperatures and chemical
potentials. We will employ $T_9$ to denote the temperature in units of
$10^9$K, and $\mu_{10}$ to represent the electron chemical potential in
units of 10 MeV. The result for the bremsstrahlung radiation flux from
Eq.~(\ref{finale}) can then be recast as
\beqy
F=\int_{z=0}^{z=z_0}dz~Q\,;\quad Q&=&1.14\times 10^{42}\biggl\{1+\frac{0.15\mu_{10}^2}{0.3\mu_{10}^2+0.26}-1.61\mu_{10}{\rm tan}^{-1}\biggl(\frac{0.92}{\mu_{10}}\biggr)\biggr\}\mu_{10}^4\biggl[1+\frac{\pi^2}{256}\biggl(\frac{T_9}{\mu_{10}}\biggr)^2\biggr]^2\\
\nonumber 
&\times&{\rm ln}(19.6\mu_{10})
\times\left\{\begin{array}{c}\frac{0.034T_9}{\mu_{10}}\biggl({\rm ln}2-0.5\biggr);~
\frac{0.15}{\mu_{10}^2}\leq\frac{T_9}{\mu_{10}}\ll 116\\0.02~{\rm e}^{-0.15/T_9\mu_{10}};~T_9/\mu_{10}\ll
0.15\end{array}\right\}{\rm erg~cm}^{-3}{\rm
s}^{-1}.\label{numemiss} 
\eeqy 
where $z_0$ is the thickness of the electrosphere. The
plasma frequency of a highly degenerate electron plasma depends
linearly on the chemical potential (see Eq.~(\ref{omegap})) and has
been rescaled accordingly ($\omega_p(z)=0.55\mu_{10}(z)$ MeV). Excepting
the outermost regions of the electrosphere, we can safely assume the
electrons to be degenerate for all chemical potentials and
temperatures considered in this work. The flux (similarly
re-expressed) from the pair creation process is~\cite{Usov01}
\beqy F_{\pm}=\int_{z=0}^{z=z_0}dz~Q_{\pm}~\,;\quad Q_{\pm}&=&4.52\times
10^{42}\{0.51+0.0862T_9\}\mu_{10}T_9^3\biggl[1+\frac{\pi^2}{256}\biggl(\frac{T_9}{\mu_{10}}\biggr)^2\biggr]^2\\
\nonumber &\times&{\rm
exp}\biggl(-\frac{11.83}{T_9}\biggr)J\biggl(\frac{11.6\mu_{10}}{T_9}\biggr)
~{\rm erg~cm}^{-3}{\rm s}^{-1}\quad,\label{numschwinger} 
\eeqy 
where
\beq 
J(x)=\frac{1}{3}\frac{x^3{\rm
ln}(1+2/x)}{(1+0.074x)^3}+\frac{\pi^5}{6}\frac{x^4}{(13.9+x)^4}\quad.
\eeq 
The blackbody flux is given by the Stefan-Boltzmann
law: 
\beq 
F_{BB}=5.67\times 10^{31}T_9^4 ~{\rm erg~cm}^{-2}{\rm s}^{-1}\,.  
\eeq 
  For illustration, we choose $T_9=1$, an electrosphere
  thickness $z_0=1000$ fm and $\mu_{10}(0)=1$.  With the electron
  density profile given by Eq.~(\ref{eprofile}), we find
\beqy 
F&=&
9.06\times 10^{28}~{\rm erg~cm}^{-2}~{\rm s}^{-1}\\ \nonumber
F_{\pm}&=& 1.45\times 10^{28}~{\rm erg~cm}^{-2}~{\rm s}^{-1}\\
\nonumber F_{BB}&=& 5.67 \times 10^{31}~{\rm erg~cm}^{-2}~{\rm
s}^{-1}\quad .  
\eeqy 
These numbers demonstrate that the bremsstrahlung radiation flux
from $e^-e^-$ scattering in the electron layer dominates over that
from pair creation. 

\begin{figure}[!ht]
\bce
\epsfig{file=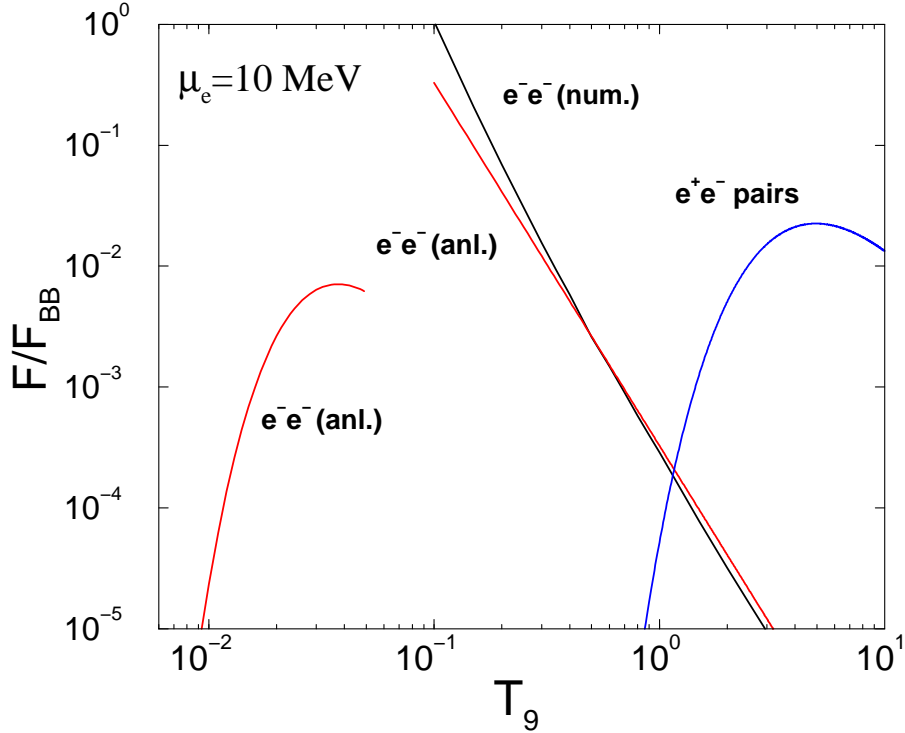,width=10.0cm,angle=270}
\ece
\caption{Photon fluxes from the bremsstrahlung and pair creation
processes scaled to the blackbody limit as a function of temperature
in units of $10^9$K ($T_9$). The curve marked (num.) shows results of
numerical integrations of Eq.~(\ref{emiss3}) which is to be compared
with the results of the analytical expressions (anl.) from
Eq.~(\ref{finale}) for low and high temperatures. The (uniform)
electron chemical potential is $\mu_e=10$ MeV with $z_0=100$ fm.}
\label{figps2}
\end{figure}

In Fig.~\ref{figps2}, we compare the flux from the bremsstrahlung
process to that from the pair creation process, both scaled to
the blackbody limit.  For simplicity and for the sake of comparison
with the analytical result Eq.~(\ref{finale}), we assume a uniform
chemical potential. In this case, the integral over $z$ gives simply
$z_0$, the thickness of the electrosphere. To be consistent, we take
the thickness to be $z_0=100$ fm for which the chemical potential
changes only by about 20$\%$ .  The analytical estimate in
Eq.~(\ref{finale}) agrees well with results from numerical
integrations of Eq.~(\ref{emiss3}). For $T_9 \leq 2$, the flux from
the bremsstrahlung process is more than that from the pair
creation process (it may even dominate at somewhat higher
temperatures once the contribution from higher energy photons is
included.) whereas for temperatures $T_9>2$, the flux from the
pair creation process exceeds that from bremsstrahlung.  (This conclusion
is not significantly modified upon the inclusion of LPM effects, as we
will see.)  Thus, for electron plasmas below a temperature of $\sim
0.1$ MeV, bremsstrahlung radiation is an important channel for photon
cooling of bare quark stars.  Its dependence on the third power of the
electron chemical potential (or linear in electron density)
distinguishes it from other processes.

With increasing electron degeneracy (value of $\mu_e/T$) or
electrosphere thickness $z_0$, the flux from bremsstrahlung
rapidly increases until it starts to become comparable to the
blackbody limit. At this stage, the inverse process, photon absorption
on electrons and rescattering (Comptonization) becomes 
important and the opacity of the layer increases. 
Detailed balance then implies that the blackbody limit is saturated,
and the electrosphere radiates like a blackbody. The inverse processes
have not been taken into account in the curves of Fig.~\ref{figps2},
hence they appear to exceed the blackbody limit for some
temperatures. However, there exists a window, $0.1<T_9<1$, in which
the inverse processes can be neglected and in which the bremsstrahlung
flux exceeds that from pair creation.

\subsection{Spectral Features and Typical Energies} 

\begin{figure}[!ht]
\bce
\epsfig{file=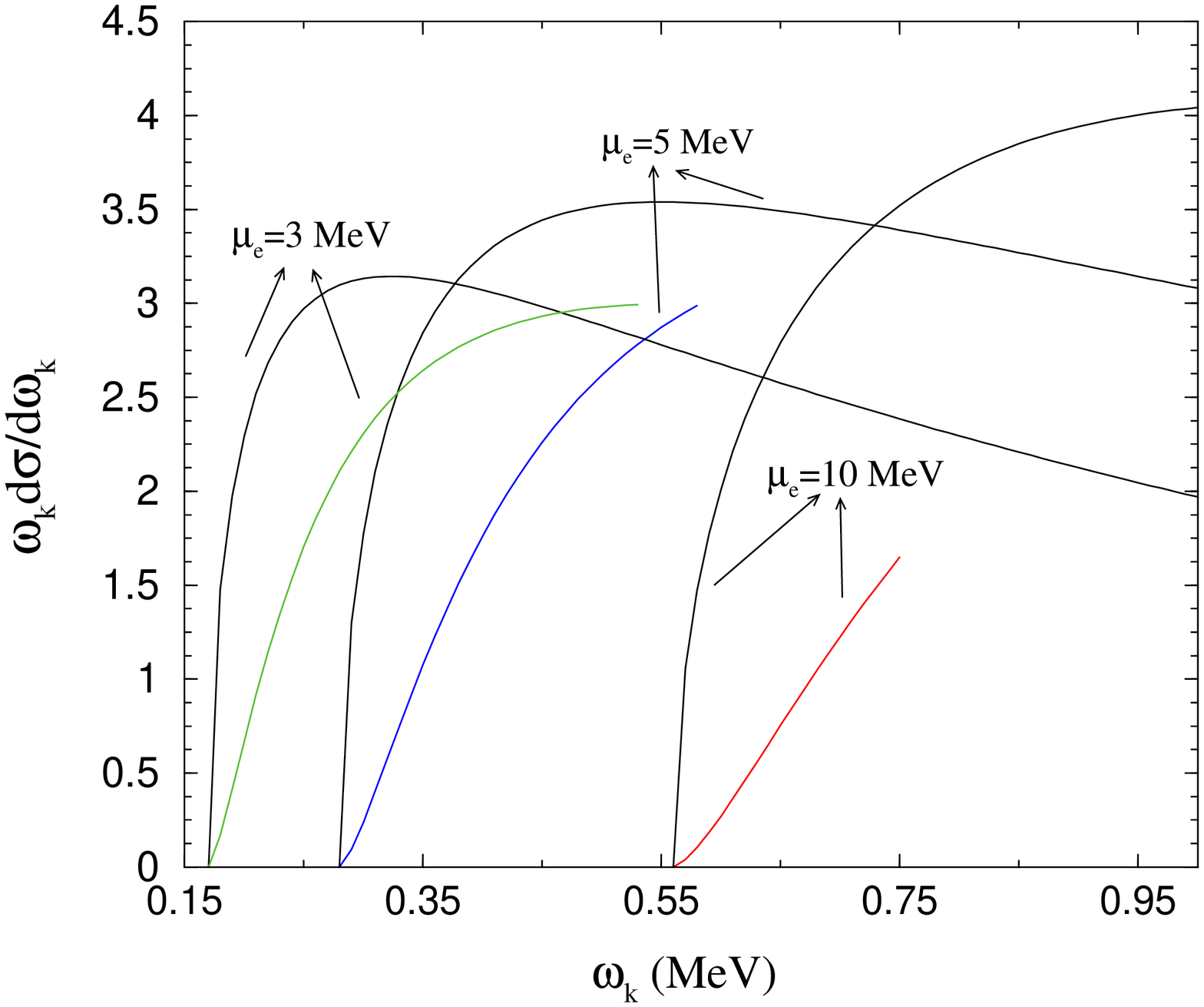,width=10.0cm,angle=270}
\ece
\caption{The energy spectrum of bremsstrahlung photons at the
indicated values of the electron chemical potential $\mu_e$. The full
curves are the spectrum from Eq.~(\ref{spectrum}), while the truncated
curves correspond to Eq.~(\ref{landaueqn}) for the
low energy spectrum in vacuum valid for frequencies satisfying the condition
$\mu_e-\omega_k\gg m_e$.}

\label{espec}
\end{figure}

The spectrum of low energy bremsstrahlung photons can be read off from
Eq.~(\ref{spectrum}). Explicitly, 
\beq
\omega_k\biggl(\frac{d\sigma}{d\omega_k}\biggr) \propto 
\biggl(1-\frac{\omega_p^2}{\omega_k^2}\biggr)^{3/2} ~~
{\rm ln}\biggl(\frac{\mu_e^2}{m_e\omega_k}\biggr)\quad, 
\eeq
where in the term involving the logarithm the electron energy $E$ has
been replaced by $\mu_e$.  In the quasiclassical approximation
employed here, support for this spectrum exists only between
$\omega_{min} = \omega_p$ and $\omega_{max} = (\omega_p^2 +
m_e^2)^{1/2}$. One must go beyond the quasiclassical approximation in
order to obtain the spectrum for $\omega > \omega_{max}$.
In vacuum, the low energy bremsstrahlung spectrum for
frequencies $\mu_e-\omega_k\gg m_e$ reads~\cite{LL}
\beq
\omega_k\biggl(\frac{d\sigma}{d\omega_k}\biggr) \propto x\left(1-\frac{\omega_p^2}{\omega_k^2}\right)^{1/2}\biggl(x+\frac{1}{x}-\frac{2}{3}\biggr)\biggl[{\rm ln}\biggl(\frac{2x\mu_e^2}{m_e\omega_k}\biggr)-\frac{1}{2}\biggr];\quad x=1-\frac{\omega_k}{\mu_e}\,.\label{landaueqn}
\eeq
For typical values of $\mu_e$ encountered across the electrosphere,
Fig.~\ref{espec} shows the role of medium effects by comparing the low
energy spectrum with the low energy spectrum neglecting all medium
effects (except the plasma cutoff).

Although photons are emitted from thermal electrons, the non-thermal
character of the bremsstrahlung radiation is evident from this
figure. While this is a natural feature of any spectrum calculated
with a low-energy cut-off, the
large mean free path of photons in the electrosphere, and medium
modifications reinforce the same physical conclusion even if the entire
range of frequencies is covered. The extra factor of $n^2$ induced by
bremsstrahlung in the medium produces a harder spectrum than in
vacuum, while the total flux is smaller. The spectrum of high energy photons
is expected to fall off steeply due to the large electron degeneracy. However,
as mentioned previously, this feature cannot be calculated within the
quasiclassical approximation made here.

We turn now to estimate the characteristic energies of the emitted
photons, and to contrast them
with those characteristic of thermal blackbody radiation. In order to
perform these tasks, we examine the quantities
\beqy
\left( \begin{array}{l} \Gamma \\ Q \\ Q^2 \\ \end{array} \right) 
&=&
\frac{1}{\hbar}
\int\frac{d^3p_1}{(2\pi)^3}\frac{d^3p_2}{(2\pi)^3}
n_F(\omega_{p_1})n_F(\omega_{p_2})
\int d\omega_k\,
\left( \begin{array}{l} 1 \\ \omega_k\\ {\omega_k}^2\\ \end{array} \right) 
\biggl(\frac{d\sigma}{d\omega_k}\biggr) \,,
\label{gqq2}
\eeqy
where $\Gamma$ is the number of bremsstrahlung reactions per unit
volume per unit time (the rate), $Q$ is the energy loss
in photons per unit volume per unit time (the emissivity), and $Q^2$
is the squared energy loss per unit volume per unit time. Good
measures of the average photon energy and the average squared photon
energy are provided  by the ratios
\beq
\langle \omega \rangle = {Q}/{\Gamma} \qquad {\rm and} \qquad  
\langle \omega^2 \rangle = {Q^2}/{\Gamma} \,.
\eeq
It is also instructive to examine the dimensionless ratio 
\beq 
\label{rat}
R = {\langle \omega^2 \rangle} / {\langle \omega \rangle^2} \,,
\eeq  
which can be used to gauge the gross spectral features
of the energy spectrum. The quantity $R$  
takes into account the full effects of the transition
amplitudes and 4-momentum conservation.

The calculations of $\Gamma$ (this is finite because of the finite
plasma frequency) and $Q^2$ for the bremsstrahlung process can be
performed following the procedure adopted for the calculation of
$Q$ in Sec. III A. The results are
\beqy
\Gamma &=&Q\frac{{\cal H}(\omega_p)}{{\cal F}(\omega_p)};\quad 
{\cal H} = \left[\frac{1}{m_e}{\rm ln}
\biggl(\frac{m_e+\sqrt{\omega_p^2+m_e^2}}{\omega_p}\biggr)-\frac{3\omega_p^2+4m_e^2}{3(\omega_p^2+m_e^2)^{3/2}}\right] \\ 
Q^2 &=& Q \frac{{\cal G}(\omega_p)}{{\cal F}(\omega_p)};\quad 
{\cal G} = \left[\frac{3\omega_p^2+m_e^2}{2\sqrt{\omega_p^2+m_e^2}}-\frac{3\omega_p^2}{2m_e}{\rm ln} 
\biggl(\frac{m_e+\sqrt{\omega_p^2+m_e^2}}{\omega_p}\biggr)\right] \,, 
\label{meanenergy}
\eeqy
where the quantities $Q$ and ${\cal F}$ are given in Eq.~(\ref{finale}). The
average photon energy is therefore 
\beqy
\langle \omega \rangle = \frac {\cal F}{\cal H} \rightarrow 
 \left\{
\begin{array}{ll}
  \omega_p & \text{for $\omega_p \gg m_e$ } \\
  m_e/\ln (2m_e/\omega_p) & \text{for $\omega_p \ll m_e$}
\end{array} \right. \,.
\eeqy
In Fig. 4, we show the mean energy of photons versus distance in the
electrosphere for $\mu_e(0)=10$ MeV and $T=1$ MeV.  The solid curve
shows the result obtained by using the analytical expression for
$\langle \omega \rangle = {\cal F}/{\cal H}$. Results obtained by
numerical integrations of Eq.~(\ref{emiss3}) and its counterpart for
$\Gamma$ are shown by the dashed curve.  It is clear that the
analytical results give a fairly accurate representation of the
numerical results for values of $\omega_p$ sampled across the
electrosphere (see Fig. 1). The mean energy of photons varies from 0.4
MeV to 0.7 MeV, depending on the depth from which the photons are
emitted. On average, this is larger than the mean energy of 0.5 MeV
from the annihilation of $e^+e^-$ pairs. This is also very different from the
spectrum of photons from a normal cooling neutron star with
$0.1<E/{\rm keV}<2.5$.

\begin{figure}[!ht]
\bce
\epsfig{file=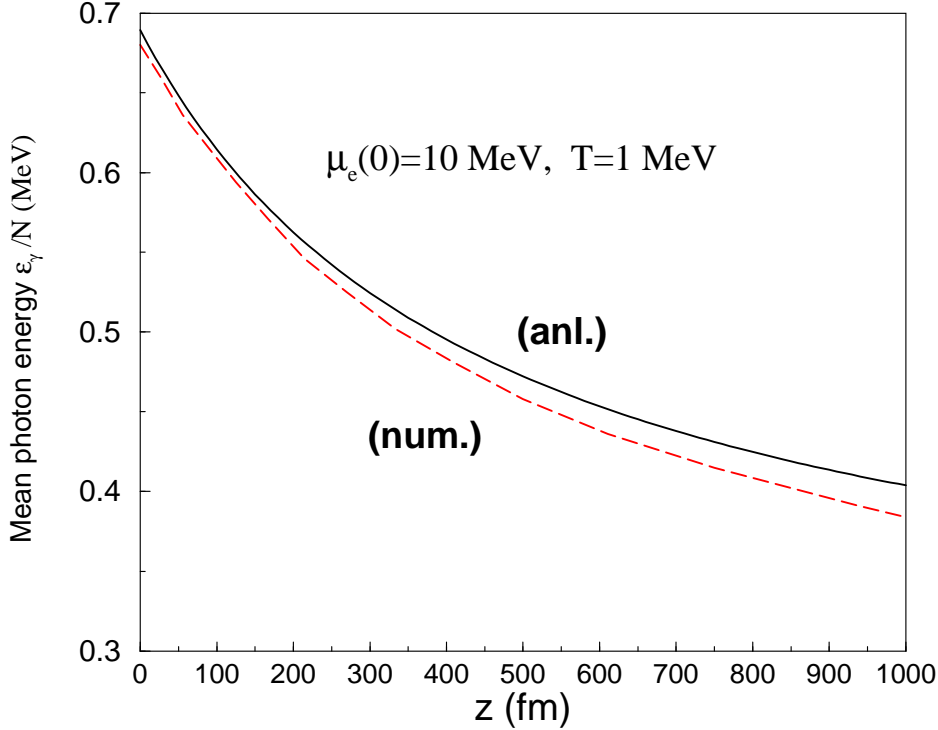,width=10.0cm,angle=270}
\ece
\caption{The mean photon energy as a function of distance in the
electrosphere. The solid curve shows results from the analytical
expression in Eq.~(\ref{meanenergy}), while the dashed curve is obtained
by numerical integrations of $Q$ in Eq.~(\ref{emiss3}) and its
counterpart $\Gamma$. Numerical values of the electron chemical
potential $\mu_e$ and temperature $T$ are as shown in the inset. }
\label{figps4}
\end{figure} 

The ratio $R$ is given by 
\beqy
\label{ratB}
R = \frac {{\cal G}{\cal H}}{{\cal F}^2} \rightarrow
 \left\{
		\begin{array}{ll}
1 & \text{for $\omega_p \gg m_e$ } \\
\frac 12\ln \left(\frac {2m_e}{\omega_p}\right) & \text{for $\omega_p \ll m_e$}
		\end{array} \right.
\eeqy

In order to highlight the differences between the bremsstrahlung and
blackbody radiation, the quantities $\langle\omega\rangle$ and $R$ can
be contrasted with the ideal gas benchmark quantities. 
\beqy 
\label{ratid}
R_{id} = \frac{{\langle \omega^2 \rangle}_{id}}
{{\langle \omega \rangle^2}_{id}} 
\qquad {\rm with} \qquad
\left( \begin{array}{l} {\langle \omega \rangle}_{id} \\
{\langle \omega^2 \rangle}_{id}  \end{array} \right) 
=
\frac{\displaystyle\int_0^\infty dk~k^2 
\left( \begin{array}{l} \omega \\ \omega^2 \\ \end{array} \right) 
F_{\gamma}(\omega)}
{\displaystyle\int_0^\infty dk \, k^2 F_{\gamma}(\omega)} \,, 
\eeqy
where $F_\gamma$ is the equilibrium Bose-Einstein distribution
function characterizing in-medium photons with mass $\omega_p$ and
zero chemical potential. Explicitly,
\beqy
{\langle\omega\rangle}_{id} 
\rightarrow
 \left\{
		\begin{array}{ll}

\omega_p + \frac 32 T 
 & \text{for $\omega_p \gg T$ } \\
 {\displaystyle{ \frac{\Gamma(4)\zeta(4)}{\Gamma(3)\zeta(3)}~T}} \simeq 2.7~T  
 & \text{for $\omega_p \ll T$ } 
\end{array} \right.
\eeqy
\beqy
{\langle \omega^2 \rangle}_{id} \rightarrow  
 \left\{
		\begin{array}{ll}
\omega_p^2 + 3mT  
 & \text{for $\omega_p \gg T$ } \\
 {\displaystyle{ \frac{\Gamma(5)\zeta(5)}{\Gamma(3)\zeta(3)}~T^2}} 
\simeq 10.35~ T^2  
 & \text{for $\omega_p \ll T$ } 
\end{array} \right.
\eeqy
These  results imply that \cite{PP03}
\beqy
R_{id} \rightarrow  
 \left\{
		\begin{array}{ll}
1   & \text{for $\omega_p \gg T$ } \\
1.42 & \text{for $\omega_p \ll T$ } 
\end{array} \right.
\eeqy

The behaviors of $R$ versus $\omega_p/m_e$ and $R_{id}$ versus
$\omega_p/T$ are compared in Fig. 5, which highlights the non-thermal
nature of the bremsstrahlung radiation emitted from degenerate
electrons. In light of the limited frequency support for
bremsstrahlung, a more rigorous quantum calculation is called for to
characterize the spectrum more accurately.

\begin{figure}[!ht]
\bce
\epsfig{file=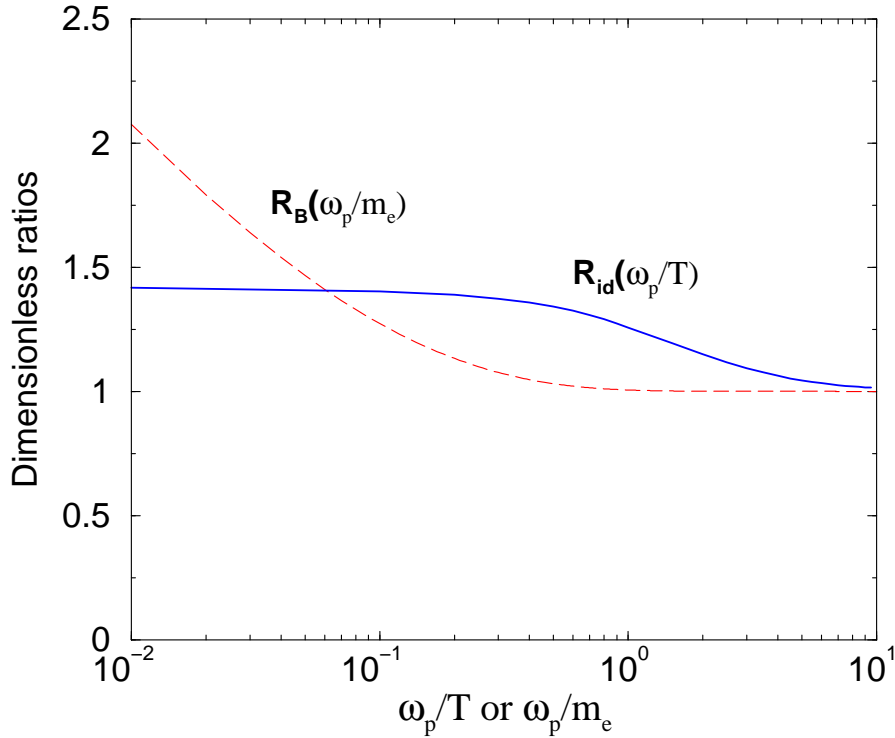,width=10.0cm,angle=270}
\ece
\caption{A comparison of the ratio $R$ in Eq.~(\ref{ratB}) 
for bremsstrahlung photons and $R_{id}$  in Eq.~(\ref{ratid}) for 
thermal photons with in-medium photon mass $\omega_p$. }
\label{figR}
\end{figure}

%-------------------------------------------------%   
\subsection{Inclusion of the Landau-Pomeranchuk-Migdal Effect}
%\subsection{Discussion}

As noted in Sec. II, multiple scattering of electrons within the
formation time of the radiated photon can lead to the so-called LPM
suppression of bremsstrahlung. We had omitted a detailed consideration
of this effect in order to isolate the $T$ and $\mu_e$-dependences of the
emissivity in the absence of the LPM effect. Since an analytical
estimate including LPM suppression is cumbersome, we have performed a
numerical integration of ~Eq.~(\ref{emiss3}) utilizing the approximate
suppression factor $S_{LPM}(k)$ from Eq.~(\ref{emiss1}). 

In Fig. 6, the effects of the LPM effect are assessed.  The LPM effect
reduces the simple analytical estimate in Eq.~(\ref{finale}) by a
significant amount. Specifically, for $\mu_{10}=1$, the reduction
factor (relative to the emissivity without the LPM effect) is $\sim
60$, while for $\mu_{10}=2$, the reduction factor is $\sim 350$. The
temperature range is $0.1 < T_9 < 10$. The suppression is stronger at
higher densities (compare the left and right panels), since the
density of scatterers controls the LPM effect. The suppression factor
for $\mu_{10}=0.5$ (not shown in the figure) is approximately a factor of 10.

\begin{figure}[!ht]
\bce
\epsfig{file=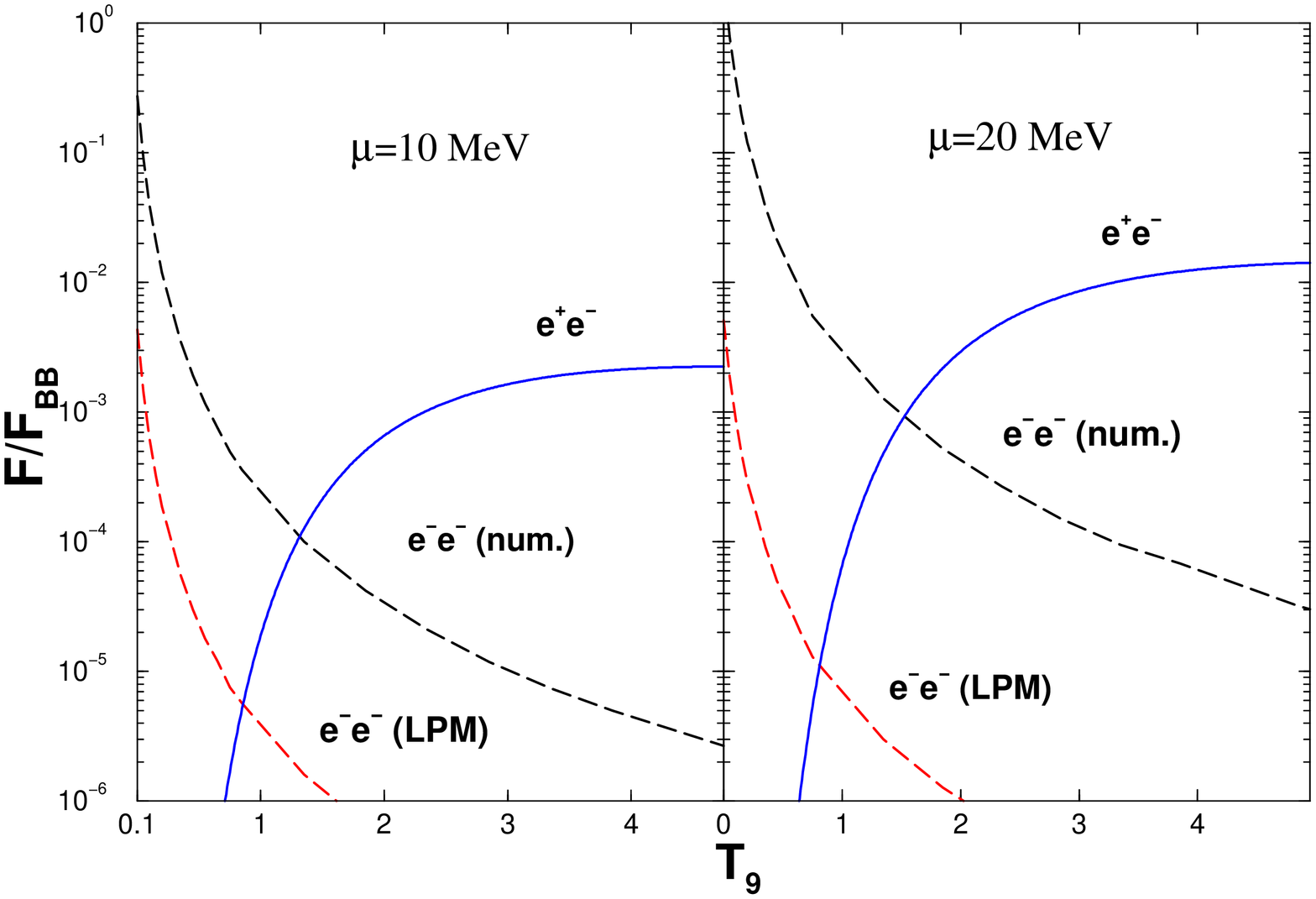,width=10.0cm,angle=270}
\ece
\caption{Photon fluxes from the bremsstrahlung and pair
creation processes scaled to the blackbody value as a
function of temperature in units of $10^9$K ($T_9$).  In the case of
the bremsstrahlung process, the upper (lower) curve shows results
without (with) the effects of the Landau-Pomeranchuk-Migdal effect.
Numerical values of the electron chemical potentials are as indicated. }
\label{figlpm}
\end{figure}

The mean photon energy $Q/\Gamma$ does not change much from the
analytical estimate upon the inclusion of the LPM suppression
factor. This is because both the rate and the energy loss are affected
similarly and therefore the LPM effect nearly cancels in the
ratio. For the same reason, the LPM suppression does not significantly
change the shape of the spectra discussed in the previous
section. The spectra are controlled primarily by the
ratio of plasma frequency to electron mass.

Equilibrium radiation from quark matter and bremsstrahlung radiation
from quark-quark collisions in the surface layers also produce
photons.  In Ref.~\cite{Chmaj91}, luminosities from these processes
were compared to the blackbody luminosity; it was found that for
$T\leq 2\times 10^{10}$ K (i.e., $T_9<20$), the equilibrium radiation
is negligible.  For these temperatures, the non-equilibrium radiation
from quark-quark Bremstrahlung dominates, although it is still 4
orders of magnitude less than the blackbody luminosity (even this may
be an overestimate according to Ref.~\cite{Usov01}). Accounting for
the multiple scattering (LPM effect) of quarks and absorption by the
electron layer leads to the conclusion that an additional suppression
by 2 orders of magnitude is likely~\cite{Cheng03}, so that overall,
the photon emissivity from quark matter is about 6 orders of magnitude
smaller than the blackbody emissivity. The dominant contribution thus
comes from $e^+e^-$ pair production via the Schwinger mechanism in the
presence of strong electric fields or the $e^-e^-$ bremsstrahlung
process in the thin electron layer.  The relative
importance of these two processes depends sensitively on the 
temperature.

In the range $T\leq 10^{9}$K, bremsstrahlung is the dominant 
emission process, since the emissivity from the pair creation
process is vanishingly small.  The analytical treatment of the
bremsstrahlung process performed in this work provides a good
approximation to the numerical results in this range of
temperatures. At low temperatures ($T \leq 10^7$ K), the
bremsstrahlung rate is exponentially small as seen from
Eq.~(\ref{finale}). For temperatures in the range $T\sim 10^8$K, the
inverse process of photon absorption also becomes important as the
rate approaches the blackbody limit.  The main conclusion is that for
temperatures $T<10^9$K, the bremsstrahlung process is likely to be the dominant
photon emission source, even in the presence of LPM effects, when compared to
the annihilation of $e^+e^-$ pairs.

\section{Conclusions and Outlook}
We have calculated the photon emissivity from the $e^-e^- \rightarrow
e^-e^-\gamma$ bremsstrahlung process occurring in the electrosphere of
a bare strange quark star. For degenerate electrons and for low energy
photons, unscreened magnetic interactions provide the largest
contribution to the emissivity, while the contribution from screened
electric interactions is comparatively negligible.  For temperatures
in the range $10^8$K$<T<10^9$K, the emissivity from the
bremsstrahlung process surpasses that of the $e^+e^-$ pair
production process occurring in the presence of a strong electric field in
the electrosphere. Within the soft photon approximation, the mean
energy of the bremsstrahlung photons escaping the electrosphere is
$\sim 0.5$ MeV. This estimate may be slightly altered
if the entire spectrum is sampled. Under strongly degenerate
conditions, the contribution of high energy photons to the emissivity
is expected to be small in comparison to that from low energy photons.

Photons from the bremsstrahlung process are non-thermal for
sufficiently thin electrospheres and moderate electron chemical
potentials, because such photons do not undergo sufficient number of
scatterings in the electrosphere.  Inclusion of the
Landau-Pomeranchuk-Migdal effect, which arises due to multiple
scatterings of the emitting electron, reveals that even in the
presence of a large suppression from this effect the bremsstrahlung
process emerges as the dominant source of radiation from the star's
surface for temperatures in the range $10^8$K$<T<10^9$K.  We
may regard the LPM-inclusive bremstrahlung rates as a conservative
estimate, since we have approximated photon exchange by a static
Coulomb field.  If retardation effects were included, the suppression
will be smaller leading to somwewhat higher rates. 
Both the ``bare'' and LPM-corrected rates take into account the
dielectric suppression effect of the electrosphere. The consequence is
that the LPM suppression is two orders of magnitude less than that in
the case the photon is emitted into vacuum. There is an additional
mitigating effect of about 15\% on the LPM suppression due to the
strong external electric field, which can bend the electron trajectory
causing the formation time of the emitted photon wave packet to
decrease.  This implies that the emission angles can be increased
beyond $\theta_{max}=m_e/E$. However, this only changes the emissivity
by a pre-factor $\sim 4/3$, and to logarithmic accuracy our principal
results are valid even in the presence of strong electric fields. In
view of this, our LPM-corrected rates should be treated as a lower
bound.

An intriguing result of our calculations is that for $T\leq
10^{-3}~\mu_e(0)$, bremsstrahlung emission becomes comparable to
blackbody emission. This indicates that bremstrahlung absorption has
to be taken into account in estimating the optical thickness of the
electrosphere (see the discussion following Eq.~(\ref{Thomson})). This
will lead to better estimates of the emissivity applicable in the
temperature window excluded by Eq.~(\ref{numemiss}).
The remaining task is to perform calculations of the thermal evolution
of strange quark matter stars including the non-thermal bremsstrahlung
and thermal pair annihilation processes in order to provide baseline
calculations of the surface luminosity versus age.
Results of such calculations will be reported separately.

\section*{Acknowledgements}
We thank Vladimir Usov for urging us to investigate the bremsstrahlung
process in degenerate matter, and Guy Moore for an observation concerning
the LPM effect. P.J. and C.G. are supported in part by
the Natural Sciences and Engineering Research Council of Canada and in
part by the Fonds Nature et Technologies of Quebec. The research of
M.P. was supported by the U.S. Department of Energy grant
DOE/DE-FG02-87ER-40317 and the NSF grant INT-9802680. The work of
D.P. is partially supported by grants from UNAM-DGAPA
(PAPIIT-IN112502) and Conacyt (36632-E). \rm

\appendix 
\section{Angular Integration}
\noindent The factor ${\cal I}(T,\mu_e)$ in Eq.~(\ref{efinal}) is
evaluated below.
Since only the relative angle between $\theta_1$ and $\theta_2$ enters
in the integration, and we integrate over all relative orientations,
we can choose a direction ${\hat p_1}={\hat z}$ to decouple the two
integrations. Then, 
\beqy
\label{App1}
&&\int d\Omega_1\int d({\rm cos}~\theta)d\phi~\frac{\tilde{n}_F({\cal
E})^2}{{\cal E}^2} \\ \nonumber 
&&=8\pi^2\int_{-1}^{1}dx\left(\frac{1}{1+{\rm
e}^{(\mu_e-[\frac{p_{F_e}^2}{2}(1-x)+m_e^2]^{1/2})/T}}\right)^2\frac{1}{\frac{p_{F_e}^2}{2}(1-x)+m_e^2} \,. 
\eeqy 
Since $p_{F_e}\gg m_e$ and $T$, the dominant contribution comes from
the region around $x\approx -1$;
elsewhere the integrand is exponentially suppressed. Therefore, we can
restrict the range of $x$ from -1 to $-1+\delta$, where $\delta$ is
small. Defining $\epsilon=(1+x)/2$, we can
integrate $\epsilon$ from 0 up to a characteristic $\delta\sim 10(T/\mu_e)$ when
the exponential suppression becomes effective. This leads to 
\beq
{\cal I}\approx
\frac{16\pi^2}{p_{F_e}^2}\int_0^{\delta}
\frac{d\epsilon}{(1-\epsilon)}\biggl(\frac{1}{1+{\rm
e}^{\mu_e\epsilon/2T}}\biggr)^2\approx \frac{32\pi^2T}{\mu_e^3}
\biggl[{\rm ln}2-\frac{1}{2}\biggr] \,.
\eeq 
This result is valid as long as $m_e^2/2\mu_e T\leq 1$ and is
accurate up to corrections of higher powers of $T/\mu_e$. Therefore, it
is a better approximation for large electron degeneracies. When $m_e^2/2\mu_e
T\gg 1$, the blocking factors approach zero due to increasing
degeneracy of the Fermi sphere, and we obtain 
\beq 
{\cal I}=\frac{16\pi^2}{\mu_e^2}{\rm e}^{-m_e^2/2\mu_e T} \,. 
\eeq 
This result shows that the emissivity is 
exponentially small at very low temperatures.

%\include{Dany2}

%-----------------------THE REFERENCES------------------------------------
\begin{flushleft}

\end{flushleft}
\end{document}